\documentclass[prl,
preprintnumbers,amsmath,amssymb]{revtex4}
\usepackage{graphicx}

\DeclareGraphicsExtensions{.pdf,*.jpg}

 \def\be{\begin{equation}}
 \def\ee{\end{equation}}
 \def\bes{\begin{eqnarray}}
 \def\ees{\end{eqnarray}}

 \def\2{\frac{1}{2}}
 \def\4{\frac{1}{4}}

\catcode`\@=11
\def\@citex[#1]#2{%
\if@filesw \immediate \write \@auxout {\string \citation {#2}}\fi
\@tempcntb\m@ne \let\@h@ld\relax \def\@citea{}%
\@cite{%
  \@for \@citeb:=#2\do {%
    \@ifundefined {b@\@citeb}%
      {\@h@ld\@citea\@tempcntb\m@ne{\bf ?}%
      \@warning {Citation `\@citeb ' on page \thepage \space
undefined}}%
      {\@tempcnta\@tempcntb \advance\@tempcnta\@ne%
      \@tempcntb\number\csname b@\@citeb \endcsname \relax%
      \ifnum\@tempcnta=\@tempcntb 
it
        \ifx\@h@ld\relax%
          \edef \@h@ld{\@citea\csname b@\@citeb\endcsname}%
        \else%
          \edef\@h@ld{\ifmmode{-}\else--\fi\csname
b@\@citeb\endcsname}%
        \fi%
      \else
        \@h@ld\@citea\csname b@\@citeb \endcsname%
        \let\@h@ld\relax%
      \fi}%
    \def\@citea{,\penalty\@highpenalty\,}%
  }\@h@ld
}{#1}}

\def\@citeb#1#2{{[#1]\if@tempswa , #2\fi}}
\def\@citeu#1#2{{$^{#1}$\if@tempswa , #2\fi }}
\def\@citep#1#2{{#1\if@tempswa , #2\fi}}

\begin{document}

\title{The BTZ black hole with a time-dependent boundary}

\author{N. Lamprou}
\author{S. Nonis}
\author{N. Tetradis}
\affiliation{%
Department of Physics,
University of Athens,
University Campus,
Zographou, 157 84, Greece}
\date{\today}%

\begin{abstract}
The non-rotating BTZ solution is expressed in terms of coordinates that allow for an arbitrary time-dependent scale factor in
the boundary metric. We provide explicit expressions for the coordinate transformation that generates this form of the metric, and
determine the regions of the complete Penrose diagram that are convered by our parametrization.
This construction is utilized in order to compute the stress-energy tensor of the dual CFT on a time-dependent background.
We study in detail the expansion of radial null geodesic congruences in the BTZ background,
for various forms of the scale factor of the boundary metric.  
We also discuss the relevance of our construction for the holographic calculation of the entanglement entropy of the dual
CFT on time-dependent backgrounds. 
 
\end{abstract}

\maketitle

\section{Introduction}

The BTZ black hole \cite{btz1,btz2} is a solution of (2+1)-dimensional gravity that shares many of the
characteristics of higher-dimensional black holes. Despite the absence of a curvature singularity at the
origin, it has an event horizon, a Hawking temperature and interesting thermodynamic properties. 
(For a review, see ref. \cite{carlip}.)  These can be matched, via the AdS/CFT correspondence \cite{adscft}, to
properties of two-dimensional conformal field theories (CFTs). (For a review, see ref. \cite{kraus}.) 
The BTZ solution is locally isometric to anti-de Sitter (AdS) space. Its nontrivial structure is 
obtained through appropriate identifications. In the parameterization of the metric through 
``Schwarzschild" coordinates, one identifies the points corresponding to the 
values $\phi$ and $\phi+2\pi$ of the angular variable.  

In the context of the AdS/CFT correspondence, the properties of the dual 
two-dimensional CFTs are usually investigated on flat backgrounds. This is achieved through 
parameterizations of the BTZ solution that result in a flat boundary, such as a torus.
In general, the boundary metric of an asymptotically AdS space belongs to a conformal class. 
Starting from a bulk metric with a flat boundary, it is possible to perform a coordinate transformation 
that generates a conformally flat boundary.   In this work we carry out this procedure for the non-rotating BTZ solution,
in order to produce boundary metrics with conformal (or scale) factors that have explicit time dependence. 
The bulk metric is cast in the ``Fefferman-Graham" form \cite{fg}, so that the application of holographic
renormalization and the calculation of the stress-energy tensor \cite{skenderis} are automatic.

This procedure was carried out for the (4+1)-dimensional AdS black hole in ref. \cite{tetradis}. (See also ref. \cite{kajantie}, as well
as ref. \cite{meyer} for a generalization with a bulk dilaton.) 
In that case the generated
boundary metric was of the Friedmann-Robertson-Walker (FRW) type. Because of the higher complexity of 
the higher-dimensional problem, the coordinate transformation was not produced in closed analytical
form. The BTZ black hole provides a simpler setting, within which explicit expressions can be derived. 
Moreover, it is possible to determine precisely the part of the geometry that is covered by the coordinate chart. 
The Penrose diagram of the non-rotating BTZ black hole is similar to that of the maximal Schwarzschild geometry \cite{carlip}. 
For a static boundary, the Fefferman-Graham coordinates cover the regions outside the two event horizons.
For a time-dependent boundary, the coordinates cover part of the interior as well. A constant-time surface
has the form of a throat, or Einstein-Rosen bridge \cite{mtw}, connecting two asymptotically AdS regions. For a time-dependent
boundary, this throat
may stretch behind the event horizon. In the main part of the paper
we investigate in detail the radial null geodesics for various types of time dependence of the 
boundary metric. We determine the expansion of both ingoing and outgoing geodesic congruences and study how they
are affected by the presence of the time-dependent throat. 

In the following section we discuss the parameterization with a static boundary. We present the transformation from
Schwarzschild to Fefferman-Graham coordinates and derive the stress-energy tensor of the dual CFT on a 
flat background. In section 3 we present the transformation that results in a boundary with a time-dependent 
scale factor. The corresponding stress-energy tensor displays the correct conformal anomaly. 
In section 4 we present the conformal diagrams of the BTZ black hole for various forms of the boundary metric. They 
offer a visual perception of the regions covered by the parametrization.
In section 5 we study the ingoing and outgoing radial null geodesics for various forms of the boundary scale factor. 
In the conclusions we discuss possible applications of our results in the context of the AdS/CFT, such as the calculation of
the entanglement entropy for CFTs on time-dependent backgrounds.

We mention at this point that the metric with a linearly expanding boundary has also been analyzed in ref. \cite{louko}. 
In this work we generalize that analysis for a boundary metric with an arbitrary scale factor. Our results are in agreement
with \cite{louko} for a linear scale factor. We also study in detail the
form of the ingoing and outgoing null geodesics and determine the nature of the various surfaces of vanishing null
expansion that appear in the Fefferman-Graham system of coordinates.

\section{Static boundary}

The non-rotating BTZ black hole \cite{btz1} in 2+1 dimensions is a solution of the Einstein field equations with a 
negative cosmological constant: 
$\Lambda_3 = -1/l^2$.
We set $l =1$ for simplicity.
The metric can be written in Schwarzschild coordinates as
\begin{equation}
\label{eqmetric} 
ds^2 = -f(r) dt^2 + \frac{dr^2}{f(r)} + r^2 d\phi^2,\ \ \ \ \ \ f(r) = r^2- \mu. 
\end{equation}
The Hawking temperature, energy (or mass) and entropy of the black hole are, respectively,
\begin{equation}
\label{eqTM}
 T = \frac{1}{2\pi}\sqrt{\mu}, 
\ \ \ \ \ \ \ 
E = \frac{V}{16\pi G_3}\mu,
\ \ \ \ \ \ \ 
S = \frac{V}{4 G_3}\sqrt{\mu},
\end{equation}
with $V$ the volume of the one-dimensional space spanned by $\phi$, 
and $G_3$ Newton's constant. In order for the above solution to have the properties of a black hole, the coordinate
$\phi$ must be periodic, with period equal to $2\pi$.

In order to discuss the AdS/CFT correspondence, it is convenient to express the metric in 
terms of Fefferman-Graham coordinates \cite{fg}. This can be achieved by defining a new coordinate $z$ through
${dz}/{z} = - {dr}/{\sqrt{f(r)}}. $
We obtain
\begin{equation}
z=\frac{2}{\mu}\left(r-\sqrt{r^2-\mu} \right),
\label{zofrm}
\end{equation}
where we have chosen the multiplicative constant so that $z\simeq 1/r$ for $r\to \infty$.
Inverting this relation we find 
\begin{equation}
r=\frac{1}{z}+\frac{\mu}{4}z.
\label{rofz}
\end{equation}
The coordinate $z$ takes values in the interval $[0,z_e=2/\sqrt{\mu}]$, covering the region outside the event
horizon, in which $r$ takes values in the interval $[r_e=\sqrt{\mu},\infty]$.
The metric of eq. (\ref{eqmetric}) takes the form
\begin{equation}
\label{eqmetric1} 
ds^2 = \frac{1}{z^2} 
\left[ dz^2 - \left( 1-\frac{\mu}{4}z^2\right)^2 dt^2 +\left( 1+\frac{\mu}{4}z^2\right)^2 d\phi^2 \right]. 
\end{equation}

It is important to notice that eq. (\ref{rofz}) is also satisfied for 
\begin{equation}
z=\frac{2}{\mu}\left(r+\sqrt{r^2-\mu} \right),
\label{zofrp}
\end{equation}
which results from the condition
${dz}/{z} = + {dr}/{\sqrt{f(r)}}. $
With this choice of sign, the metric of eq. (\ref{eqmetric}) takes again the form
of eq. (\ref{eqmetric1}). However, the coordinate $z$ now takes values 
in the interval $[z_e=2/\sqrt{\mu},\infty]$, covering the region outside the event
horizon, in which $r$ takes values in the interval $[r_e=\sqrt{\mu},\infty]$.

We may consider the metric of eq. (\ref{eqmetric1}), allowing $z$ to vary in the interval $[0,\infty]$.
In this case the region outside the horizon is covered twice. 
The coordinates $(t,z)$ are closely related to the isotropic coordinates that are 
often employed for the study of the Schwarzschild geometry.
The isotropic coordinates do not span the full space. 
They cover the two regions of the Kruskal-Szekeres plane that are located
outside the horizons.   
The same happens for the coordinates $(t,z)$ in the case of the BTZ black hole in 2+1 dimensions, or
the AdS-Schwazschild geometry in higher dimensions. For fixed coordinate time $t$, the metric describes a throat, or
Einstein-Rosen bridge, connecting two asymptotically flat regions \cite{mtw}. In the following we shall
generalize this construction in a time-dependent setting.

The Hawking temperature $T$ 
of the black hole \cite{page} can be
determined from the metric of eq. (\ref{eqmetric1}). In Euclidean space, 
the metric possesses a conical singularity at the location of the horizon
$z_e= 2/\sqrt{\mu}$. This can be eliminated if the Euclidean time is 
periodic, with period $1/T$. Expanding $z$ around $z_e$, we find that the conical
singularity disappears for $T$ taking the value 
given by the first of eqs. (\ref{eqTM}).

The energy and entropy of the black hole can be related 
to those of the dual CFT on the AdS boundary. The most general (2+1)-dimensional metric that satisfies Einstein's equations with 
a negative cosmological constant is of the form \cite{finite}
\be\label{eq2} ds^2 = \frac{1}{z^2} \left[ dz^2 + g_{\mu\nu} dx^\mu dx^\nu \right], \ee
where
\be g_{\mu\nu} = g_{\mu\nu}^{(0)} + z^2 g_{\mu\nu}^{(2)} 
+ z^4 g_{\mu\nu}^{(4)}.  \ee
(However, the global properties of the geometry can be quite non-trivial \cite{global}.)
The stress-energy tensor of the dual CFT is \cite{skenderis}
\begin{equation}
\label{eq3a} 
\langle T_{\mu\nu}^{(CFT)} \rangle =
\frac{1}{8\pi G_3} 
\left[ g^{(2)}-{\rm tr}\left( g^{(2)}\right)g^{(0)} \right].
\end{equation}
Applying this general expression to our metric (\ref{eqmetric1}), 
we obtain the energy density and pressure, respectively,
\begin{eqnarray}
\label{eq3e} 
\rho&=&\frac{E}{V}=-\langle T_{~t}^{t\,(CFT)} \rangle=\frac{\mu}{16\pi G_3},
\\
P& =& \langle T_{~x}^{x\,(CFT)} \rangle = 
\frac{\mu}{16\pi G_3},
\label{eq3p}  
\end{eqnarray}
on a boundary with metric 
\begin{equation}
\label{eqmetric0} 
ds_0^2 = g_{\mu\nu}^{(0)} dx^\mu dx^\nu = -dt^2 + d\phi^2.
\end{equation}

In order to determine the entropy of the CFT, we consider a variation of the 
parameter $\mu$ of the metric (\ref{eqmetric1}). This variation does not 
affect the volume $V$ of the boundary. The variations of the energy density 
$E$ and entropy $S$
of the CFT obey $dE=TdS$. A simple integration returns the expression for the entropy given in the third 
of eqs. (\ref{eqTM}).
As expected, the entropy is proportional to the surface of the event horizon.

\section{Time-dependent boundary}

We are interested in generalizing the previous discussion to the case 
of a time-dependent boundary of the form
\begin{equation}
\label{eqmetricb} ds_0^2 = g_{\mu\nu}^{(0)} dx^\mu dx^\nu = 
-d\tau^2 + a^2(\tau) d\phi^2.
\end{equation}
Through a redefinition of the time coordinate, the scale factor $a(\tau)$ can be turned into a conformal factor 
of the boundary metric.
In order to apply holographic renormalization, 
we use a foliation  
consisting of hypersurfaces whose metric is asymptotically proportional to (\ref{eqmetricb}).
The form of eq. (\ref{eqmetric1}) suggests the {\em ansatz}
\begin{equation}
\label{eqmetric33} 
ds^2
= \frac{1}{z^2} \left[ dz^2 
- \left(1-\frac{A_1(\tau)}{4}z^2 \right)^2 d\tau^2 + a^2(\tau)\left(1+\frac{A_2(\tau)}{4}z^2 \right)^2  d\phi^2 \right].
\end{equation}
The $\tau z$ Einstein equation imposes
\begin{equation}
\label{eintz}
A_1=-A_2-\frac{a\dot{A}_2}{\dot{a}},
\end{equation}
where the dot denotes a derivative with respect to $\tau$.
Substitution of this relation in the $zz$ Einstein equation results in
\begin{equation}
\label{einzz}
2a\dot{a}A_2+a^2\dot{A}_2+2\dot{a}\ddot{a}=0,
\end{equation}
which is solved by 
$A_2(\tau)=(\mu-\dot{a}^2)/{a^2}$.
We have chosen the integration constant so that the static metric of eq. (\ref{eqmetric1}) is reproduced for $a(\tau)=1$.
Finally, from eq. (\ref{eintz}) we find 
$A_1(\tau)=(\mu-\dot{a}^2+2 a \ddot{a})/{a^2}$.
In summary, we have  
\begin{equation}
\label{eqmetric3} 
ds^2= \frac{1}{z^2} \left[ dz^2 
- \mathcal{N}^2(\tau,z) d\tau^2 + \mathcal{A}^2 (\tau,z) d\phi^2 \right],
\end{equation}
with
\begin{eqnarray}
\label{aa}
\mathcal{A}(\tau,z) &=& a\left(1+ \frac{\mu-\dot{a}^2}{4a^2}z^2\right) \\
\label{nn}
\mathcal{N}(\tau,z) &=&1- \frac{\mu-\dot{a}^2+2 a \ddot{a}}{4a^2}z^2
=\frac{\dot{\mathcal{A}}(\tau,z)}{\dot{a}}.
\end{eqnarray}

The metric (\ref{eqmetric3}) is a generalization of the static metric discussed in the previous section. The static
form (\ref{eqmetric1}) is reproduced for $a(\tau)=1$.
The comparison of eqs. (\ref{eqmetric1}), (\ref{eqmetric3})
indicates that 
\begin{equation}
 r = \frac{\mathcal{A}}{z}= \frac{a}{z} +\frac{\mu-\dot{a}^2}{4} \frac{z}{a}. 
\label{eqsys4} \end{equation}
The coordinates $(\tau,z)$ do not span the full BTZ geometry. As in the static case, they cover the two regions outside the 
event horizons,
located at
\begin{eqnarray}
\label{ze1}
 z_{e1}&=&\frac{2a}{\sqrt{\mu}+\dot{a}},
\\
\label{ze2}
z_{e2}&=&\frac{2a}{\sqrt{\mu}-\dot{a}}.
\end{eqnarray}
The quantities $z_{e1}$, $z_{e2}$ are the two roots of the equation
$r(\tau,z)=r_e=\sqrt{\mu}$. 
Moreover, the parameterization covers part of the regions behind the horizons. For constant $\tau$, the minimal value of $r$ is obtained
for $\partial r/\partial z=0$, corresponding to 
\begin{equation}
\label{zm}
z_m=\frac{2a}{\sqrt{\mu-\dot{a}^2}},
~~~~~~~~~
r_m=\sqrt{\mu-\dot{a}^2}.
\end{equation}
Clearly, $r_m\leq r_e$. Another value of interest is the one for which
$\mathcal{N}=0$. It corresponds to the point with 
\begin{equation}
\label{za}
z_a=\frac{2a}{\sqrt{\mu-\dot{a}^2+2a \ddot{a}}},
~~~~~~~~~
r_a=\frac{\mu-\dot{a}^2+a\ddot{a}}{\sqrt{\mu-\dot{a}^2+2a\ddot{a}}}.
\end{equation}
For $\dot{a}^2, a\ddot{a}\ll \mu$, we have 
$r_m/r_e=1+{\cal O}(\mu^{-1})$, $r_a/r_e=1+{\cal O}(\mu^{-1})$, $r_m/r_a=1+{\cal O}(\mu^{-2})$.

The metric (\ref{eqmetric3}) has been derived as a solution of the Einstein equations,
without reference to the BTZ metric of eq. (\ref{eqmetric}).
The two metrics 
agree provided that 
\begin{eqnarray}
\label{eqsys1} 
\frac{(r')^2}{f(r)} - f(r) (t')^2 &=& \frac{1}{z^2} \\
\label{eqsys22}
\frac{r' \dot r}{f(r)} - f(r) t'\dot t &=& 0 \\
\frac{\dot r^2}{f(r)} - f(r) \dot t^2 &=& - \frac{\mathcal{N}^2}{z^{2}}
=-\left( \frac{1}{z} -\frac{\mu-\dot{a}^2+2 a \ddot{a}}{4a^2} z \right)^2 ,
\label{eqsys3}
\end{eqnarray}
where a prime denotes a partial derivative with respect to $z$.
The coordinate $r$ is expressed in terms of $t$, $z$ through eq. (\ref{eqsys4}).
Two of the three equations above may 
then be used to determine the derivatives $\dot t$ and $t'$. 
We obtain
\begin{equation}
\label{eqsys2} 
\dot t = -\epsilon \frac{\dot{\mathcal{A}} r'}{f \dot a}=-\epsilon\frac{{\cal N}r'}{f}, 
\ \ \ \ \ \ \ \ 
t' = -\epsilon \frac{\dot a}{zf},
\end{equation}
with $\epsilon=\pm 1$.
In fact, these expressions satisfy all three equations.  
One can verify the consistency of the system (\ref{eqsys1})-(\ref{eqsys3}) by calculating the 
mixed derivative $\dot t'$ using each of the two equations (\ref{eqsys2}), 
and showing that the two expressions match. 
For $z <z_{e1}$,
or for $z > z_{e2}$,
the integration of eqs. (\ref{eqsys1}) gives
\begin{equation}
\label{tautz1}
t(\tau,z)=\frac{\epsilon}{2\sqrt{\mu}}\log\left[
\frac{4 a^2-\left(\sqrt{\mu}+\dot{a}\right)^2 z^2 }{ 4 a^2-\left(\sqrt{\mu}-\dot{a}\right)^2 z^2}
\right]+\epsilon\, c(\tau),
\end{equation}
where the function $c(\tau)$ satisfies $\dot{c}=1/a(\tau)$. 
For $z=0$ we have $t=c$, with $c$ the conformal
time on the boundary.
For $z_{e1}<z<z_{e2}$, we find
\begin{equation}
\label{tautz2}
t(\tau,z)=\frac{\epsilon}{2\sqrt{\mu}}\log\left[
\frac{-4 a^2+\left(\sqrt{\mu}+\dot{a}\right)^2 z^2 }{ 4 a^2-\left(\sqrt{\mu}-\dot{a}\right)^2 z^2}
\right]+\epsilon \, c(\tau),
\end{equation}
The values of $\epsilon$ may be chosen differently in the regions 
$z<z_{e1}$, $z_{e1}<z<z_{e2}$, $z>z_{e2}$.
 It is obvious from eq. (\ref{tautz1}), (\ref{tautz2}) that the transformation from
$(t,r)$ to $(\tau,z)$ coordinates is always singular on the event horizons.

The observer employing the $(\tau,z)$ system of coordinates
is moving relative to the one employing the ($t,r)$ coordinates. Events occurring 
at the same value of $z$ at various times take place at locations $r={\cal A}(\tau,z)/z$. 
Events with $z\simeq 0$, which take place near the boundary, occur at $r\simeq a(\tau)/z$ in the
$(t,r)$ frame. For growing $a(\tau)$ they take place at a value of the $r$ coordinate which increases with time. 
This means that the viewpoint of the observer employing the $(\tau,z)$ system of coordinates
is that of somebody moving away from the black hole with a velocity determined by $\dot{a}(\tau)$.

\begin{figure}[t]
\includegraphics[width=100mm,height=96mm]{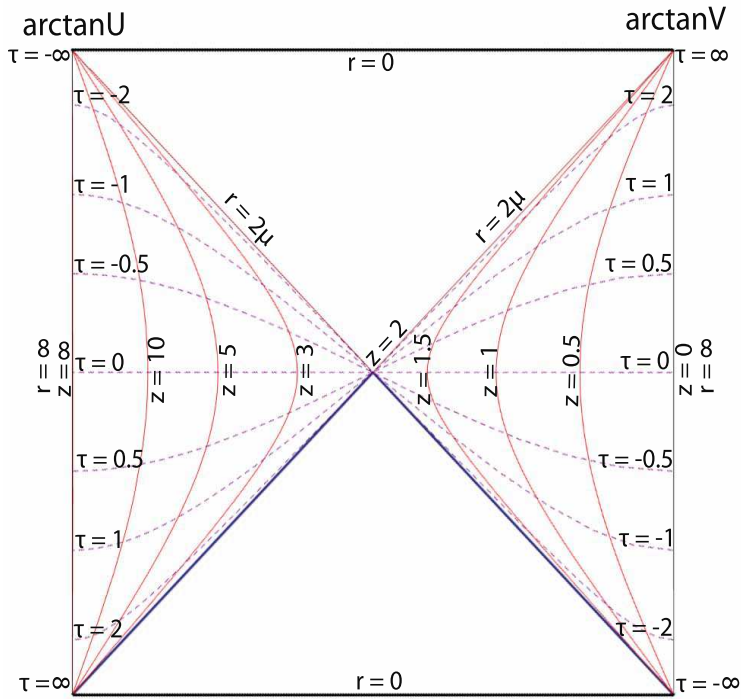}
\caption{Conformal diagram of the BTZ black hole, displaying the regions covered by the system of coordinates of eq. (\ref{eqmetric1})
with $\mu=1$. }
\label{a1}
\end{figure}

The stress-energy tensor of the dual CFT on the time-dependent boundary 
is determined via holographic renormalization and is given by eq.~(\ref{eq3a}). 
We obtain the energy density and pressure, respectively,
\begin{eqnarray}
\label{eq3te} 
\rho&=&\frac{E}{V}=-\langle T_{~t}^{t\,(CFT)} \rangle=
\frac{1}{16\pi G_3}\frac{\mu-\dot{a}^2}{a^2}
\\
P& =& \langle T_{~x}^{x\,(CFT)} \rangle = 
\frac{1}{16\pi G_3}\frac{\mu-\dot{a}^2+2a\ddot{a}}{a^2},
\label{eq3tp}  
\end{eqnarray}
on a boundary with metric 
\begin{equation}
\label{eqmetrict} 
ds_0^2 = g_{\mu\nu}^{(0)} dx^\mu dx^\nu = -dt^2 +a^2(\tau) d\phi^2.
\end{equation}
We deduce the conformal anomaly
\begin{equation}
 \langle T_{~\mu}^{\mu\,(CFT)} \rangle  = 
\frac{1}{8\pi G_3}\frac{\ddot{a}}{a}.
\label{confan}\end{equation}
These expressions agree with the ones obtained for the conformal vacuum \cite{bd}, when the 
stress-energy tensor in the Minkowski vacuum is given by eqs. (\ref{eq3e}), (\ref{eq3p}).

\section{Conformal diagrams}

\begin{figure}
\includegraphics[width=100mm,height=96mm]{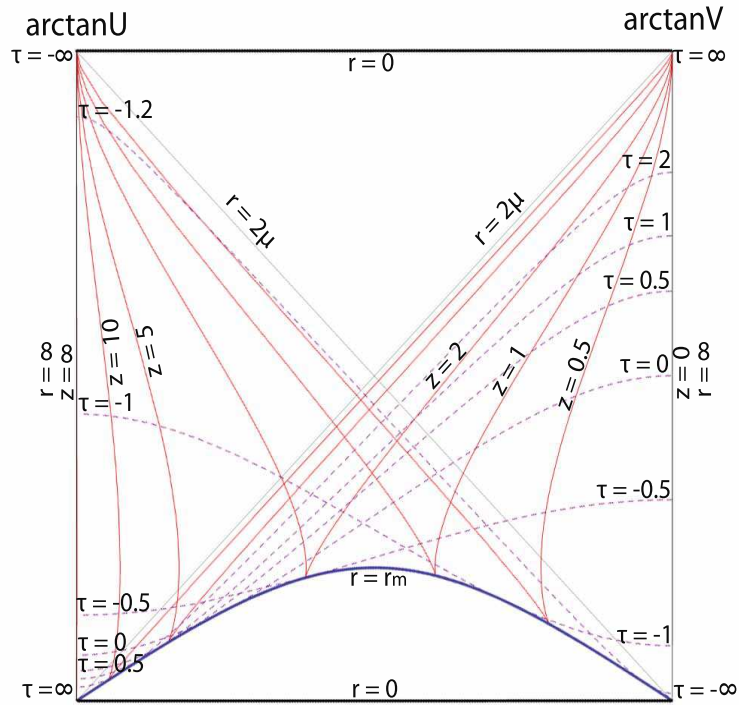}
\caption{Conformal diagram of the BTZ black hole, displaying the regions covered by the system of coordinates of eq. (\ref{eqmetric3}),
with $a(\tau)=1+v\tau$ and $\mu=1$, $v=0.8$. }
\label{v0.8}
\end{figure}

\begin{figure}
\includegraphics[width=100mm,height=96mm]{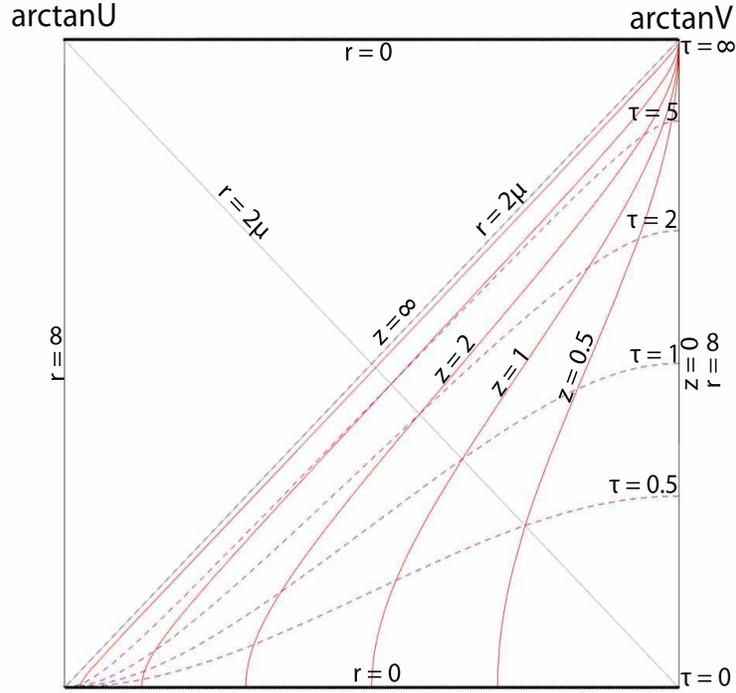}
\caption{Conformal diagram of the BTZ black hole, displaying the regions covered by the system of coordinates of eq. (\ref{eqmetric3}),
with $a(\tau)=1+v\tau$ and $\mu=1$, $v=1$.}
\label{v1}
\end{figure}

In order to provide a better understanding of the $(\tau,z)$ system of coordinates, 
we present in figs. \ref{a1}-\ref{v1}
the conformal diagram of the BTZ black hole, depicting the regions covered by the $(\tau,z)$ coordinates for a boundary metric
with a scale factor $a(\tau)=1+v \tau$. 
The coordinates ($U$,$V$) are related to $(\tau,z)$ through the relations 
\begin{eqnarray}
U&=&-\frac{2+\left(v-\sqrt{\mu} \right)z+2v \tau}{2+\left(v+\sqrt{\mu} \right)z+2v \tau}\left(1+v\tau \right)^{-\sqrt{\mu}/v}
\label{UU} \\
V&=&\frac{2-\left(v+\sqrt{\mu} \right)z+2v \tau}{2-\left(v-\sqrt{\mu} \right)z+2v \tau}\left(1+v\tau \right)^{\sqrt{\mu}/v}.
\label{VV}
\end{eqnarray}
These result from the standard relations between the ($U,V$) coordinates and the Schwarzschild coordinates $(t,r)$, and the
transformations (\ref{eqsys4}),(\ref{tautz1}),(\ref{tautz2}) with $\epsilon=1$. 

In fig. \ref{a1} we depict the part of the conformal diagram covered by the $(\tau,z)$ coordinates when $v=0$ and $a=1$.
This corresponds to the case of a static boundary. The $(\tau,z)$ coordinates 
cover the two regions outside the event horizons. In fig. \ref{v0.8} the boundary is time-dependent with
$v=0.8 < \sqrt{\mu}=1$. The $(\tau,z)$ parametrization covers the two regions outside the horizons. It also covers twice
part of a region inside the horizons, delimited by the line of minimal distance $r_m$ of eq. (\ref{zm}). Finally, fig. \ref{v1}
depicts the situation for a time-dependent boundary with $v=\sqrt{\mu}=1.$ The $(\tau,z)$ coordinates cover completely
one copy of each region inside and outside the horizons. 
The picture for $v>\sqrt{\mu}$ is similar to fig. \ref{v1}. 

We are interested in the interpretation of the geometry by an observer located on the boundary. Our motivation 
stems from our wish to interpret this construction in the context of the AdS/CFT correspondence. For this reason, in the following 
we analyze the form of ingoing and outgoing null geodesics, as well as the various surfaces on which their expansion vanishes,
within the $(\tau,z)$ system of coordinates.

\section{Null geodesics and horizons}

We would like to determine the horizons of the metric (\ref{eqmetric3}).
A possible horizon is the surface $z_a(\tau)$ for which
$\mathcal{N}(\tau,z_a)=0$. Another possibility is the event horizon 
of the static black hole, expressed in $(\tau,z)$ coordinates. 
In order to determine the nature of these surfaces, we study the 
expansion of null geodesics in the $(\tau,z)$ system of coordinates.
As can be deduced from eqs. (\ref{tautz1}), (\ref{tautz2}),
the time-coordinate $\tau$ has a non-monotonic, at points singular, dependence on the $(t,r)$ coordinates. For this 
reason, a direct comparison of the form of null geodesics in the two
coordinate systems is not straightforward. We carry out the analysis in ($\tau,z)$ coordinates, which provide the 
natural setting for an observer located at the boundary.

An apparent horizon is defined as the boundary of the region of
trapped surfaces. This boundary is a surface on which 
the expansion of outgoing null geodesics vanishes.
We denote the two sets of geodesics as the solutions of 
$(dz(\tau)/d\tau)_\pm=\mp\mathcal{N}(\tau,z)$. 
In the region near the boundary at $z=0$, where $\mathcal{N}(\tau,z_a)\simeq 1$, the solutions $z_+(\tau)$
and $z_- (\tau)$ clearly correspond to outgoing and ingoing null geodesics, respectively.  In other regions, and especially 
behind the horizons, a more careful analysis
is necessary in order to determine their nature. Quite often we refer to the geodesics $z_\pm(\tau)$ as out/ingoing, for simplicity.  
This characterization is consistent in the asymptotic regions outside the horizons. On the other hand, the true nature of the geodesics
behind the horizons becomes clear through the analysis in the rest of the paper.

The functions $z_\pm(\tau)$
define surfaces of areal radii $A(\tau,z_\pm(\tau))/z_\pm(\tau)=r_\pm(\tau)$. 
The growth of the 
volume of such a surface is proportional to the total time derivative of $r$ along 
the light path, i.e. to 
\be \label{growth}
\left(\frac{dr}{d\tau}\right)_\pm=\dot{r}+r'\left(\frac{dz}{d\tau}\right)_\pm
=\mathcal{N}(\tau,z)\left( \frac{\dot{a}}{z} \mp r' \right).
\ee 
The product 
\begin{equation}
\left(\frac{dr}{d\tau}\right)_+\left( \frac{dr}{d\tau}\right)_-=\mathcal{N}^2 
\left[ \left(\frac{\dot{a}}{z}\right)^2 -\left( r' \right)^2\right]
=-\frac{\left(\mu-\dot{a}^2\right)^2}{16a^2 z^4}\left(z^2-z_{e1}^2\right)\left(z^2-z_{e2}^2 \right)\mathcal{N}^2 
\label{hori} \end{equation}
 is expected to be invariant under
relabelling of the scalars that define null hypersurfaces \cite{kinoshita}.
This product vanishes on the event horizons, located at $z=z_{e1}(\tau)$ and $z=z_{e2}(\tau)$.
It also vanishes on the 
surface $z=z_a(\tau)$, for which $\mathcal{N}=0$. 
In the static case with $a(\tau)=1$, in which $z_{e1}=z_{e2}=z_a=2/\sqrt{\mu}= z_e$, we have
\begin{equation}
\left(\frac{dr}{d\tau}\right)_+\left(\frac{dr}{d\tau}\right)_-
=-\frac{1}{z^4}\left(z^2-z_{e}^2\right)^4.
\label{horis} \end{equation}

In order to have a clearer understanding of the nature of these surfaces, it is useful to have explicit
analytical expressions for the null geodesics. This can be achieved in specific cases.
In the following we determine the explicit form of the ingoing and outgoing radial null geodesics that start from 
a specific point $z=z_0$ at time $\tau=0$, for various forms of the function $a(\tau)$. 

\section{Specific cases}

\begin{figure}[t]
\begin{minipage}{78mm}
\includegraphics[width=\linewidth,height=78mm]{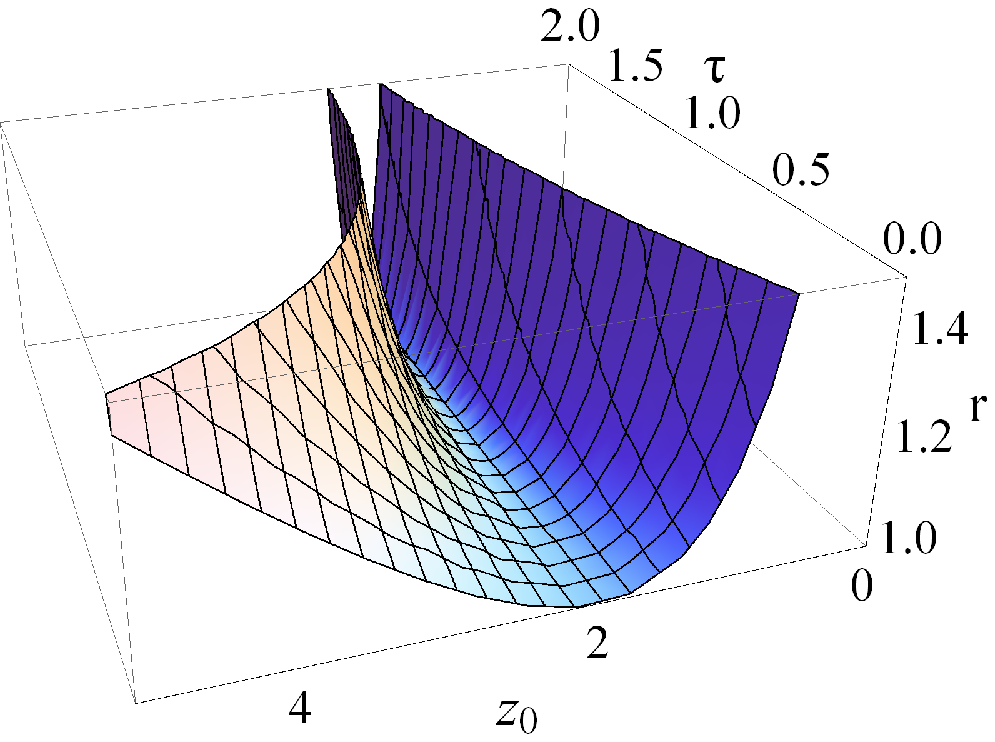}
\caption{Outgoing geodesics in the case of a static boundary.}
\label{outs}
\end{minipage}
\hfil
\begin{minipage}{78mm}
\includegraphics[width=\linewidth,height=78mm]{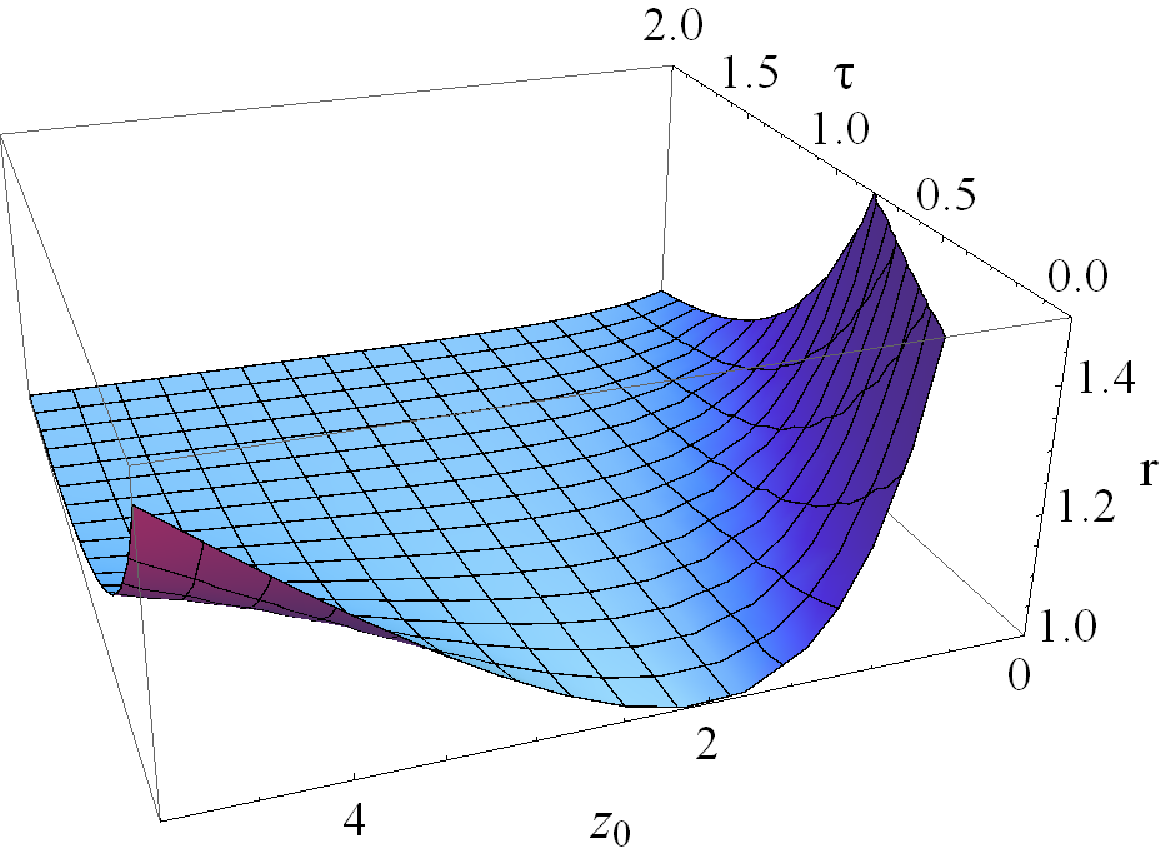}
\caption{Ingoing geodesics in the case of a static boundary.}
\label{ins}
\end{minipage}
\end{figure}

\subsection{Case I: $a=1$}

In the case of static boundary with $a(\tau)=1$, the solutions are 
\begin{eqnarray}
z_+(\tau)&=&\frac{2}{\sqrt{\mu}}\,
\frac{2+\sqrt{\mu}\,z_0+\exp\left( \sqrt{\mu}t\right) \left(-2
+\sqrt{\mu}\,z_0\right)}{2+\sqrt{\mu}\,z_0-\exp\left( \sqrt{\mu}t\right) \left(-2+\sqrt{\mu}\,z_0\right)}
\label{stout} \\
z_-(\tau)&=&\frac{2}{\sqrt{\mu}}\,
\frac{-2+\sqrt{\mu}\,z_0+\exp\left( \sqrt{\mu}t\right) \left(2
+\sqrt{\mu}\,z_0\right)}{2-\sqrt{\mu}\,z_0+\exp\left( \sqrt{\mu}t\right) \left(2+\sqrt{\mu}\,z_0\right)}.
\label{stin} 
\end{eqnarray}
The form of outgoing and ingoing geodesics is depicted in figs. \ref{outs} and \ref{ins}, respectively, for 
$\mu=1$ and for a range of emission points $z_0$. We plot the areal distance $r$ as a function of $\tau$.
In fig. \ref{outs} we observe that $r$ grows for geodesics which start on either side of the location of the 
event horizon $z_{e}=2$. For $z_e \not= 2$, light rays reach the boundary within a finite time $\tau$. Only rays which start
exactly at $z=2$ remain on the horizon with constant $r(\tau)=r_e=1$. 
The opposite behavior is observed in fig. \ref{ins}. All geodesics, irrespectively of the initial value $z_0$,
converge towards the value $r_e=1$. 

\begin{figure}[t]
\begin{minipage}{78mm}
\includegraphics[width=\linewidth,height=78mm]{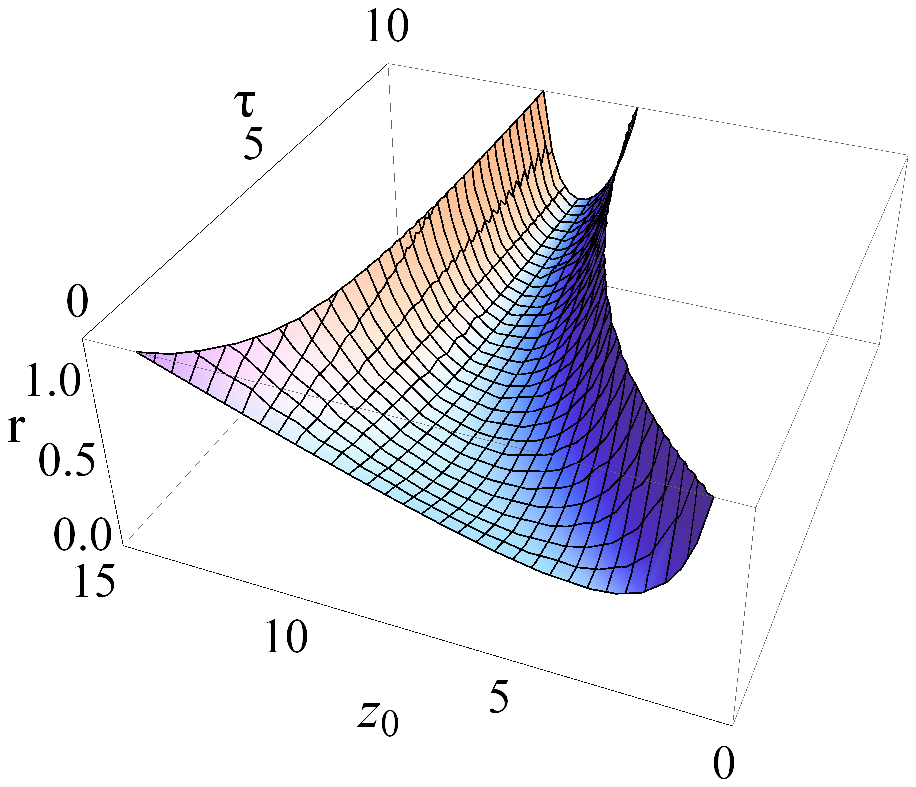}
\caption{Outgoing geodesics in the case of a time-dependent boundary with constant $\dot{a}<\sqrt{\mu}$.}
\label{outt}
\end{minipage}
\hfil
\begin{minipage}{85mm}
\includegraphics[width=\linewidth,height=78mm]{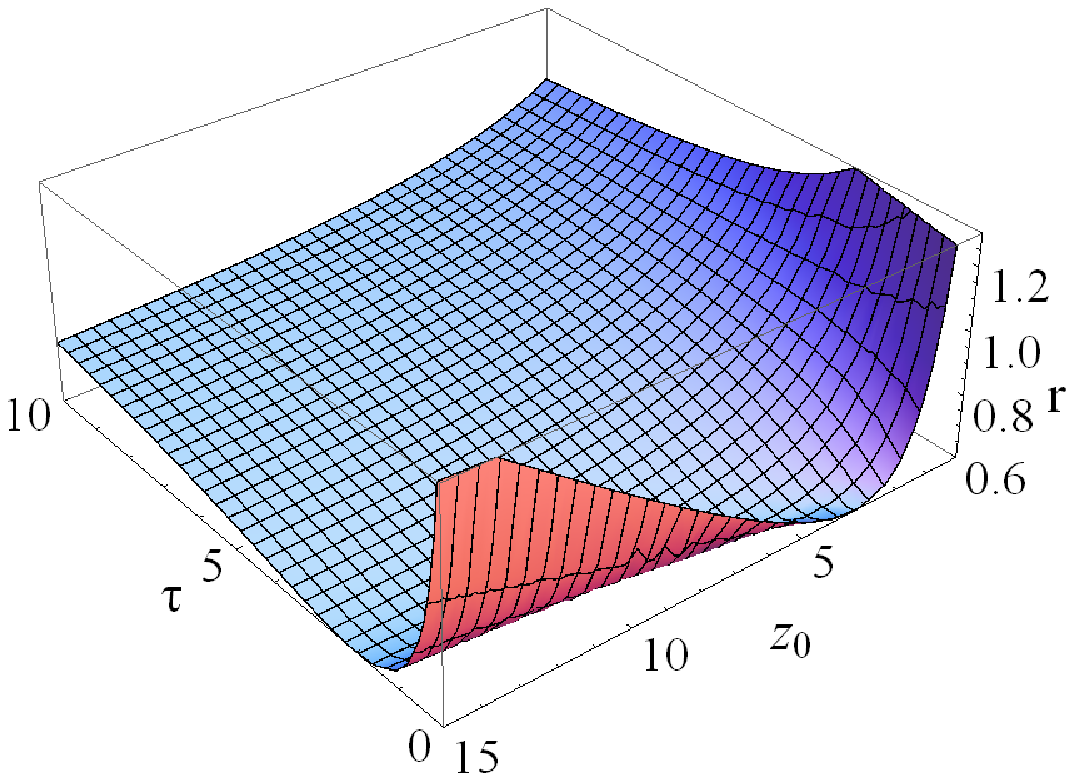}
\caption{Ingoing geodesics in the case of a time-dependent boundary with constant $\dot{a}<\sqrt{\mu}$.}
\label{int}
\end{minipage}
\end{figure}

\subsection{Case II: $a=1+v\tau$}

For a time-depended boundary the form of the null geodesics is more complicated. However, it is possible to derive
analytical solutions for the simplified case $a(\tau)=1+v \tau$, in which
the boundary metric takes the form of the Milne Universe. 
We find 
\begin{eqnarray}
z_+(\tau)&=&2(1+vt)
\frac{ 2+\sqrt{\mu}\,z_0+vz_0+(1+vt)^{\sqrt{\mu}/v}\left(-2+\sqrt{\mu}\,z_0
-vz_0 \right)}{2\sqrt{\mu}+\mu\,z_0-v(2+vz_0)-(\sqrt{\mu}+v)(1+vt)^{\sqrt{\mu}/v}\left(-2+\sqrt{\mu}\,z_0-vz_0 \right)}
\label{tout} \\
z_-(\tau)&=&2(1+vt)
\frac{ 2-\sqrt{\mu}\,z_0-vz_0+(1+vt)^{\sqrt{\mu}/v}\left(-2-\sqrt{\mu}\,z_0
+vz_0 \right)}{-2\sqrt{\mu}+\mu\,z_0+v(2-vz_0)+(\sqrt{\mu}+v)(1+vt)^{\sqrt{\mu}/v}\left(-2-\sqrt{\mu}\,z_0+vz_0 \right)}.
\label{tin} 
\end{eqnarray}

\subsubsection{$\mu>v^2$}

The form of $z_+(\tau)$ is depicted in fig. \ref{outt} for $\mu=1$ and $v=0.8$.  The presence of the second event
horizon is apparent in this plot. Its 
location is marked by the null geodesic with starting point $z_0=z_{e2}(\tau=0)=2/(\sqrt{\mu}-v)=10$, or $r_0=1$. This geodesic is 
characterized by constant areal distance $r(\tau)=r_e=1$. Geodesics with $z_0>10$ have growing areal distance, which
diverges at a finite value of $\tau$. They are outgoing null geodesics. Geodesics with $10>z_0>z_m(\tau=0)=2/\sqrt{\mu-v^2}\simeq3.33$ 
initially have $dr/d\tau <0$ and are ingoing ones. Within a finite time interval, they approach a point at which $dr/d\tau$ vanishes,
while subsequently it becomes positive. This indicates that the geodesics become outgoing. The turning point corresponds to
$r=r_m=\sqrt{\mu-v^2}=0.6$ for every geodesic.  Eventually $r(\tau)$ becomes larger than $r_e=1$, which indicates that the
geodesics cross outside the first event horizon. Geodesics with  $0.33>z>z_{e1}(\tau=0)=2/(\sqrt{\mu}+v)\simeq 1.11$ are outgoing
ones that start behind the first event horizon and eventually cross it.  
Geodesics with $z<1.11$ are outgoing ones that start outside the
first event horizon.

The form of $z_-(\tau)$ is depicted in fig. \ref{int} for $\mu=1$ and $v=0.8$. All geodesics have $r(\tau)\to 1$ for $\tau \to \infty$ in this
case. The first event horizon is apparent in this plot, marked by
the geodesic with  $z_0=z_{e1}(\tau=0)=2/(\sqrt{\mu}+v)\simeq1.11$, or $r_0=1$. This geodesic has constant  $r(\tau)=r_e=1$.
Geodesics with $z_0<1.11$ are ingoing ones, and asymptotically approach the first horizon from outside. 
Geodesics with $1.11<z_0<z_m(\tau=0)=2/\sqrt{\mu-v^2}\simeq3.33$ are outgoing ones that asymptotically approach the first horizon from inside.
Geodesics with  $3.33<z_0<z_{e2}(\tau=0)=2/(\sqrt{\mu}-v)=10$ are ingoing ones starting inside the second horizon, while those 
with $z_0>10$ are ingoing ones that start outside the second horizon and cross inside it. Both these sets of geodesics turn
outgoing at a point with $r=r_m=\sqrt{\mu-v^2}=0.6$ and asymptotically approach the first horizon from inside.

\begin{figure}[t]
\includegraphics[width=90mm,height=60mm]{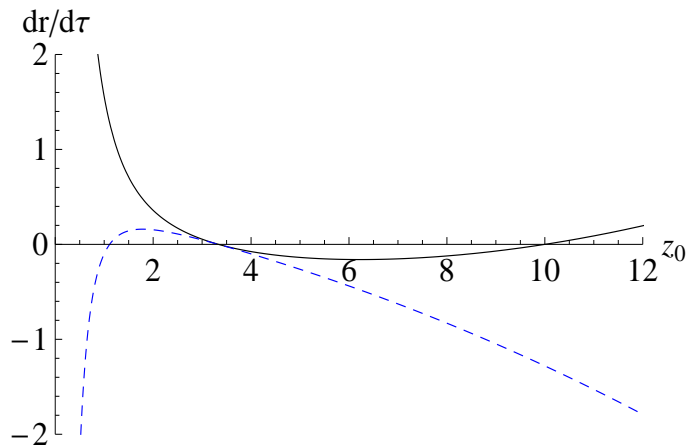}
\caption{Initial time-derivative of the areal distance for various null geodesics.}
\label{all}
\end{figure}

In order to summarize the behavior of the geodesics 
we plot in fig. \ref{all} the time-derivative $dr/d\tau$ of the areal distance at the initial time $\tau=0$, as 
a function of the value of the starting point $z_0$. This quantity determines the expansion of the
null congruence and characterizes the geodesic as ingoing or outgoing. The upper, solid curve corresponds to $(dr/d\tau)_+$, 
while the lower,
dashed one to $(dr/d\tau)_-$. At the location of the first event horizon ($z=z_{e1}(\tau=0)\simeq1.11$), the expansion of the outgoing
geodesics remains positive, while that of the ingoing ones changes sign. This behavior is identical to that encountered for the standard 
four-dimensional Schwarzschild metric in outgoing Eddington-Finkelstein coordinates. The outgoing
null geodesics cross the horizon, while the horizon is made up from world lines of
ingoing photons \cite{mtw}.
The opposite behavior is observed on the second horizon ($z=z_{e2}(\tau=0)$=10). The ingoing geodesics retain a negative
expansion, while that of the outgoing geodesics changes sign.  
In this case the event horizon is formed by outgoing photons, while the incoming
ones cross it. This is identical to what happens in the  
four-dimensional Schwarzschild geometry in ingoing Eddington-Finkelstein coordinates.
This behavior is a result of the transformation 
from $(t,r)$ to $(\tau,z)$, given by eqs. (\ref{eqsys4}), (\ref{tautz1}),  (\ref{tautz2}), which is singular on the horizon. 
It would take an infinite time $t$ for any geodesic to cross a horizon.

The expansion of all geodesics changes sign at the point of closest approach $z=z_m(\tau=0)\simeq 3.33$. This corresponds to 
$r=r_m=0.6$ and is located behind the event horizons, for which $r=r_e=1$. 
An important point is that both sets of geodesics are outgoing or ingoing on either side of 
$z_m(\tau=0)$. As a result the quantity  $(dr/d\tau)_+ (dr/d\tau)_-$, defined in eq. (\ref{hori}), remains positive behind the horizons,
even though it vanishes exactly at $z_0=z_m(\tau=0)$. This should be contrasted with what happens on the event horizons, where 
 $(dr/d\tau)_+ (dr/d\tau)_-$ changes sign. This difference indicates that, despite the fact that the expansion vanishes at
$z_0=z_m(\tau=0)$ (or $r=r_m$), this point cannot be considered as a horizon. 

As a final remark, we point out that there are no null geodesics that connect the two asymptotic regions outside the event
horizons. Ingoing geodesics that go through a horizon get trapped in the region behind the other one. Similarly, outgoing
geodesics that exit through a horizon originate in the region behind the other one. As a result, there is no communication
between the two asymptotic regions. This is consistent with what is expected for the Rosen-Einstein bridge in the
case of the Schwarzschild geometry \cite{mtw}.

\subsubsection{$\mu=v^2$}
For $a=1+v\tau$ and $\mu=v^2$ the metric (\ref{eqmetric3}) simplifies considerably and takes the form
\begin{equation}
\label{eqmetricvm} 
ds^2
= \frac{1}{z^2} \left[ dz^2 
-d\tau^2 + (1+\sqrt{\mu}\tau)^2  d\phi^2 \right].
\end{equation}
The outgoing/ingoing geodesics are given by 
\begin{equation}
z_\pm(\tau)=z_0\mp \tau,
~~~~~~~~~~~~~~
r_\pm(\tau)=\frac{1+\sqrt{\mu}\tau}{z_0\mp \tau}.
\label{geotv} 
\end{equation}

The second horizon, located at $z_{e2}$ given by eq. (\ref{ze2}), moves to infinity in this case. 
The same holds for the point of closest approach to the black hole, denoted by $z_m$ and given by eq. (\ref{zm}).
The first horizon, located at $z_{e1}$ given by eq. (\ref{ze1}), remains at a finite value. The event 
horizon corresponds to the geodesic that starts at $z_0=z_{e1}(\tau=0)=1/\sqrt{\mu}$ and has constant $r(\tau)=\sqrt{\mu}$.
The new parameterization, apart from the asymptotic region, covers the whole region behind
the horizon as well. For $\tau=0$, the center of the black hole is located at $z_0=\infty$, which corresponds to $r=0$.
The range $[1/\sqrt{\mu},\infty]$ of $z_0$ covers the whole region behind the horizon, while the range $[0,1/\sqrt{\mu}]$ 
the whole region outside
the horizon. The second copy of the black-hole in the full Penrose diagram 
does not appear in this parameterization.

Geodesics $z_-(\tau)$ with $z_0<1/\sqrt{\mu}$ approach the horizon asymptotically from outside, while the ones with 
$z_0>1/\sqrt{\mu}$ approach it from inside. The outgoing geodesics $z_+(\tau)$ are characterized by an areal distance that grows with
time and diverges at a finite value of $\tau$. The ones that start behind the horizon cross it within finite time $\tau$. 
As we mentioned earlier, this behavior is identical to that encountered for the standard 
four-dimensional Schwarzschild metric in outgoing Eddington-Finkelstein coordinates. The outgoing
null geodesics cross the horizon, while the horizon is made up from world lines of
ingoing photons \cite{mtw}.

\subsubsection{$\mu<v^2$}

For $a=1+v\tau$ and $\mu<v^2$ the geodesics are again given by eqs. (\ref{tout}), (\ref{tin}).
Their form is very similar to that in the case with $\mu=v^2$. The
only difference is that, at $\tau=0$,  the whole region behind the horizon is now covered by the range 
$[z_i,z_{e1}(\tau=0)]$ of $z_0$, with
$z_{e1}(\tau=0)=1/\sqrt{\mu}$ the location of the horizon, and 
$z_i=2/\sqrt{v^2-\mu}$ the value corresponding to $r=0$. The region outside the
horizon is covered by the range $[0,z_{e1}(\tau=0)]$ of $z_0$.

\begin{figure}[t]
\begin{minipage}{72mm}
\includegraphics[width=\linewidth,height=60mm]{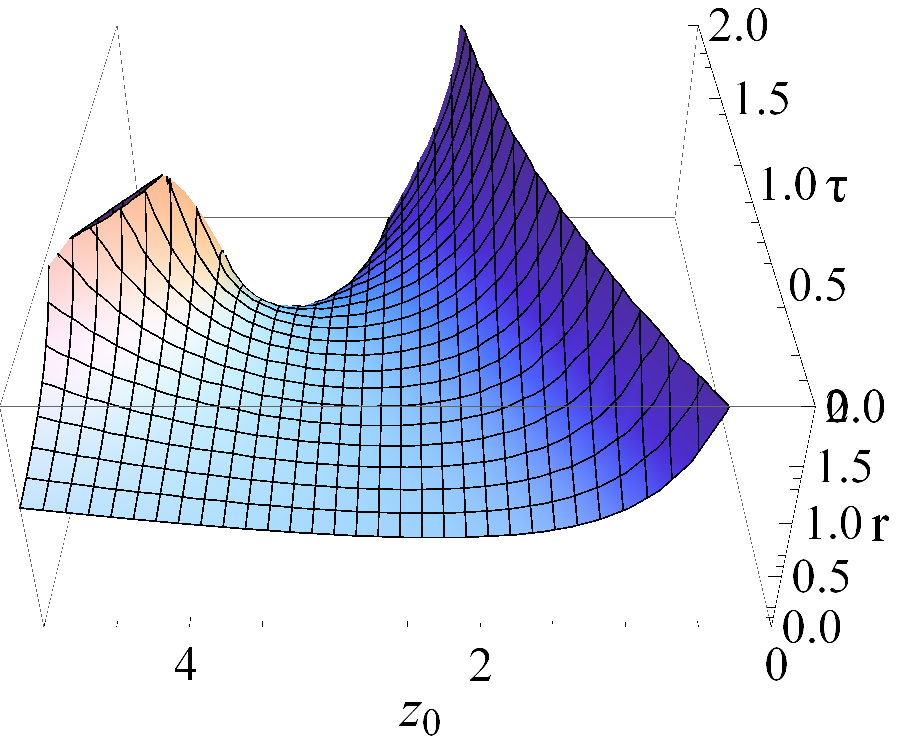}
\caption{Outgoing geodesics for a de Sitter boundary with $H<\sqrt{\mu}$.}
\label{outds1}
\end{minipage}
\hfil
\begin{minipage}{75mm}
\includegraphics[width=\linewidth,height=60mm]{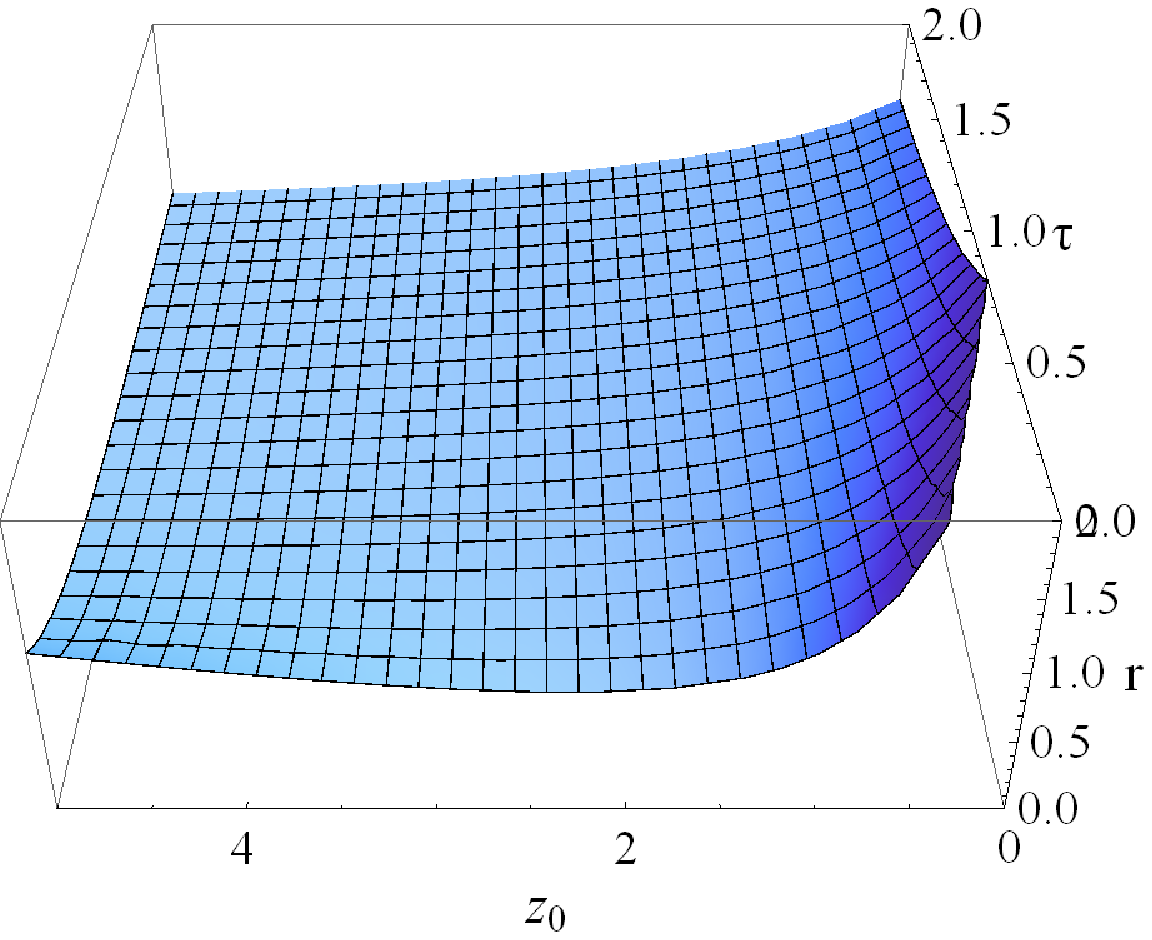}
\caption{Ingoing geodesics for a de Sitter boundary with $H<\sqrt{\mu}$.}
\label{inds1}
\end{minipage}
\end{figure}

\subsection{Case III: $a=\exp(H\tau)$} 

It is interesting to study the form of the metric when $\dot{a}$ is not constant. As a typical example we consider
the case $a=\exp(H\tau)$, in which the boundary metric has the de Sitter form.
It is possible to derive analytical expressions for the null geodesics. They are 
\begin{eqnarray}
z_+(\tau)&=& \frac{1}{2}\left[
-H+\frac{ \sqrt{\mu}\exp(-H\tau) \Bigl[ \exp\left[\sqrt{\mu}\exp(-H\tau)/H \right] 
\left(2+Hz_0+\sqrt{\mu}z_0\right) +\exp(\sqrt{\mu}/H) \left(2+Hz_0-\sqrt{\mu}z_0\right) 
\Bigr]}{ \exp\left[\sqrt{\mu}\exp(-H\tau)/H \right] 
\left(2+Hz_0+\sqrt{\mu}z_0\right) +\exp(\sqrt{\mu}/H) \left(-2-Hz_0+\sqrt{\mu}z_0\right) }
\right]^{-1}
\label{dsout} \\
z_-(\tau)&=&\frac{1}{2}\left[
H+\frac{ \sqrt{\mu}\exp(-H\tau) \Bigl[ \exp\left[\sqrt{\mu}\exp(-H\tau)/H \right] 
\left(2-Hz_0-\sqrt{\mu}z_0\right) +\exp(\sqrt{\mu}/H) \left(2-Hz_0+\sqrt{\mu}z_0\right) 
\Bigr]}{ \exp\left[\sqrt{\mu}\exp(-H\tau)/H \right] 
\left(-2+Hz_0+\sqrt{\mu}z_0\right) +\exp(\sqrt{\mu}/H) \left(2-Hz_0+\sqrt{\mu}z_0\right) }
\right]^{-1}.
\label{dsin} 
\end{eqnarray}
The event horizons of eqs. (\ref{ze1}), (\ref{ze2}) are located at
\begin{eqnarray}
z_{e1}&=&\frac{2\exp(H\tau)}{\exp(H\tau)H+\sqrt{\mu}}
\label{ze1ds}\\
z_{e2}&=&\frac{2\exp(H\tau)}{-\exp(H\tau)H+\sqrt{\mu}}.
\label{ze2ds}
\end{eqnarray}
Both these expressions correspond to $r=r_e=\sqrt{\mu}$.

\begin{figure}[t]
\begin{minipage}{72mm}
\includegraphics[width=\linewidth,height=60mm]{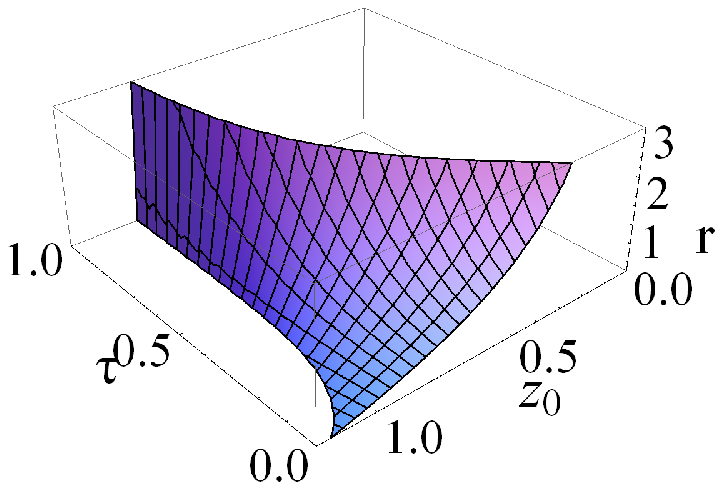}
\caption{Outgoing geodesics for a de Sitter boundary with $H>\sqrt{\mu}$.}
\label{outds2}
\end{minipage}
\hfil
\begin{minipage}{75mm}
\includegraphics[width=\linewidth,height=60mm]{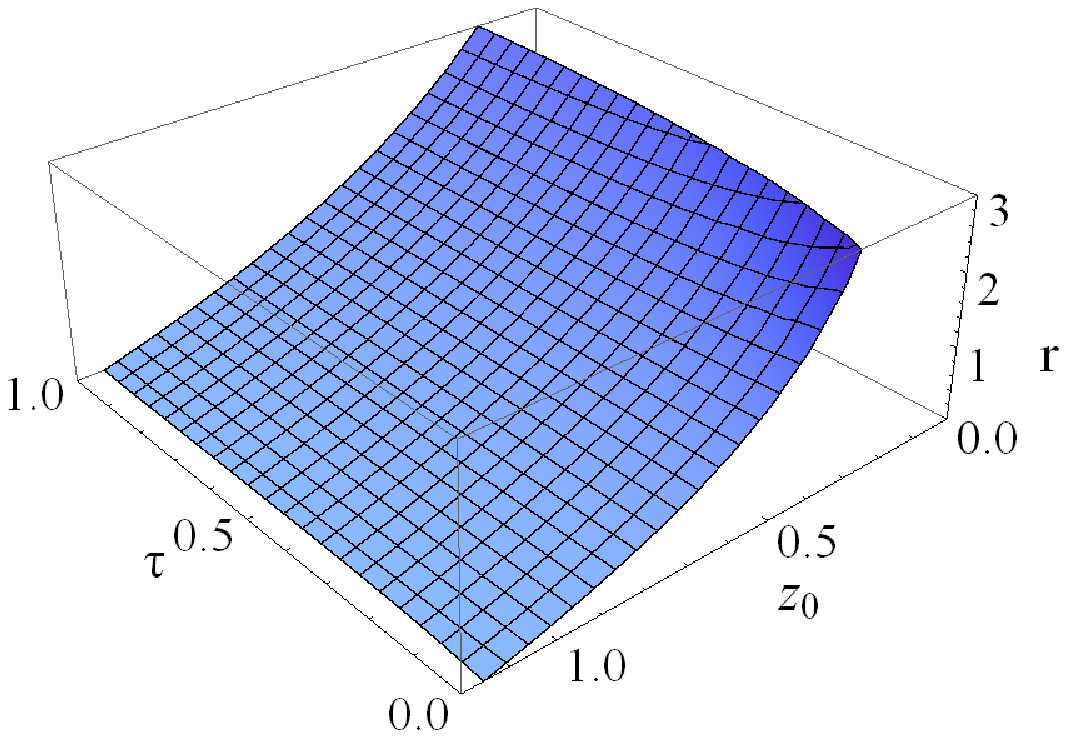}
\caption{Ingoing geodesics for a de Sitter boundary with $H>\sqrt{\mu}$.}
\label{inds2}
\end{minipage}
\end{figure}

For $\tau\ll 1/H$ the horizons are located at $z_{e1}^<=2/(H+\sqrt{\mu})$ and $z_{e2}^<=2/(-H+\sqrt{\mu})$.
For $H^2>\mu$ the second horizon does not exist, as its location corresponds to negative values of 
$z$, which are not included in the parameterization. For $\tau\gg 1/H$ the location of the first horizon is at
$z_{e1}^>=2/H$. During the time evolution the second horizon starts from $z_{e2}^<$ and moves to 
$z=\infty$ at some finite value $\tau_e=\log(\mu/H^2)/(2H)$. The first horizon shifts from $z_{e1}^<$ to 
$z_{e1}^>$ during the evolution. The point of closest approach to the black hole is located at 
\begin{equation}
\label{zmds}
z_m=\frac{2\exp(H\tau)}{\sqrt{-\exp(2H\tau)H^2+\mu}},
~~~~~~~~~
r_m=\sqrt{-\exp(2H\tau)H^2+\mu}.
\end{equation}
For $\tau \ll 1/H$, we have $z_m=z_m^<=2/\sqrt{-H^2+\mu}$ for $H^2<\mu$.  
The value of $z_m$ becomes infinite within a finite time $\tau_m=\tau_e$. At this time $r_m=0$. 
For $H^2>\mu$ the whole region behind the event horizon, with $r$ taking values in the interval $[0,\sqrt{\mu}]$, 
is covered by a finite interval of $z$. For $\tau=0$ this interval is $[2/(H+\sqrt{\mu}),2/\sqrt{H^2-\mu}]$.

\subsubsection{$H^2<\mu$}

The form of the outgoing and ingoing null geodesics for $H=0.5$, $\mu=1$ is depicted in figs. \ref{outds1}, \ref{inds1}, 
respectively. For $\tau\ll 1/H$ the behavior is similar to that depicted in figs. \ref{outt}, \ref{int}. 
There is a horizon at $z_{e2}(\tau=0)= 2/(-H+\sqrt{\mu})=4$, visible in the form of outgoing geodesics in fig. \ref{outds1}.
There is another horizon at  $z_{e1}(\tau=0)= 2/(H+\sqrt{\mu})\simeq 1.33$, visible in the form of the ingoing geodesics 
in fig. \ref{inds1}. There is also a point of
closest approach $z_m(\tau=0)=2/\sqrt{-H^2+\mu}\simeq 2.31$. The difference with figs. \ref{outt}, \ref{int} emerges 
for $\tau\gg 1/H$. In fig. \ref{outds1}, 
the geodesics with initial values $z_0$ in the interval $[2.31,4]$, which start behind the second horizon,
do not emerge through the first one at late times (as in fig. \ref{outt}). Instead, they end up at the center of the 
black hole ($r=0$). Outgoing geodesics with initial values $z_0$ in the interval $[1.33,2.31]$ do emerge as outgoing 
geodesics through the first horizon. The ingoing geodesics depicted in fig. \ref{inds1} display a different feature. 
For $\tau\to\infty$, they do not approach a horizon at $r_e=\sqrt{\mu}=1$, but some $z_0$-dependent value of $r$. 
This feature becomes more prominent for $H^2>\mu$.

\subsubsection{$H^2>\mu$}

The form of the outgoing and ingoing null geodesics for $H=2$, $\mu=1$ is depicted in figs. 
\ref{outds2}, \ref{inds2}, respectively. The whole region behind the horizon is covered by 
$z_0$ taking values in the interval $[2/(H+\sqrt{\mu}),2/\sqrt{H^2-\mu}]\simeq [0.67,1.15]$.
There is only one horizon, appearing in fig. \ref{inds2} 
at $z_{e1}(\tau=0)= 2/(H+\sqrt{\mu})\simeq 0.67$. Some of the geodesics depicted in fig. \ref{outds2} that start behind the horizon
are in reality ingoing and end up at the center of the black hole ($r=0$). 
All the outgoing geodesics that travel out of the horizon eventually reach the boundary. On the other hand, the 
ingoing geodesics of fig. \ref{inds2} do not reach the horizon for $\tau \to \infty$, but settle down at a $z_0$-dependent value
of $r$.  

\subsubsection{Asymptotic areal distance of incoming geodesics}

\begin{figure}[t]
\includegraphics[width=90mm,height=60mm]{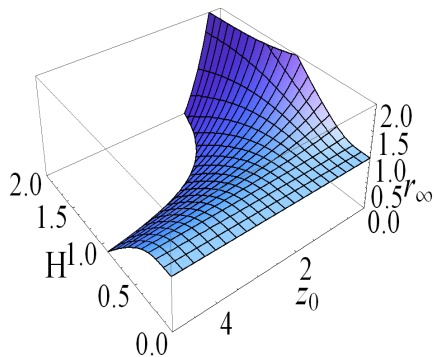}
\caption{Areal distance of incoming geodesics for $\tau\to\infty$ for a de Sitter boundary.}
\label{rinfinity}
\end{figure}

The asymptotic value of $r$ for incoming geodesics for $\tau\to\infty$ is 
\begin{equation}
r_\infty=\frac{\sqrt{\mu}\left[2-Hz_0-\sqrt{\mu}\,z_0+\exp\left(\sqrt{\mu}/H\right)\left(2-Hz_0+\sqrt{\mu}\,z_0 \right)\right]}{-2+Hz_0+\sqrt{\mu}\,z_0+\exp\left( \sqrt{\mu}/H\right)\left(2-Hz_0+\sqrt{\mu} \,z_0 \right)}
\label{rasym}
\end{equation}
In fig. \ref{rinfinity} we depict this quantity for $\mu=1$. We observe that for $H\ll \mu$ all incoming geodesics approach the
horizon located at $r_e=\sqrt{\mu}=1$. For larger values of of $H$,  geodesics that start sufficiently close to the center of the black hole
$(r=0)$ eventually reach it. The rest settle down at some value of $r$ different from 0 or 1.
In order to understand this behavior we recall our remark that near the boundary 
the observer employing the $(\tau,z)$ system of coordinates
is moving away from the black hole with a velocity determined by $\dot{a}$. An incoming geodesic has growing $z_-(\tau)$. On the
other hand this growth maybe compensated by the rapid increase of $a(\tau)$, so that, in the $(t,r)$ system of coordinates, $r=a(\tau)/z_-(\tau)$ becomes 
constant at late times. This is realized in the case in hand.

\subsection{Other cases}
In addition to the cases we considered, it is possible to derive analytical expressions for other forms of the scale function $a(\tau)$. 
One such example is the case $a(\tau)=(1+w\tau)^2$. We shall not discuss explicitly other examples. In general, 
the form of the geodesics combines characteristics from the cases we discussed above. 

\section{Conclusions}

The main aim of this work was to express the BTZ solution in a form that  contains a nontrivial boundary metric. In particular, we
were interested in boundaries analogous to the four-dimensional FRW space-times. We have shown that it is possible to achieve this 
goal through a coordinate transformation of the three-dimensional BTZ metric. The underlying reason is that, in asymptotically AdS   
spaces, the boundary metric belongs to a conformal class. It is possible to switch between members of the class through appropriate
bulk coordinate transformations. We provided explicit expressions for the transformation that turns the boundary metric from flat
to one with an arbitrary time-dependent scale factor. Moreover, the asymptotic form of the bulk metric is such that the
implementation of holographic renormalization is automatic \cite{fg,skenderis}.  It is possible to derive the renormalized 
stress-energy tensor of the two-dimensional CFT on the time-dependent background. It displays the expected properties,
including the correct conformal anomaly. 

The motivation for this study was to prepare the ground for the use of holographic methods in the computation of 
the entropy of the dual CFTs in time-dependent settings. 
The proposal of ref. \cite{ryu,takayanagi,hubeny} advocates that the entanglement entropy of a CFT within a certain region 
on the boundary can be identified with the area of an extremal  surface defined by the boundary region and extending into the
bulk. In a time-dependent situation, this surface is characterized by zero expansion of the null geodesics perpendicular to it. 
Possible (event or apparent) horizons of the bulk geometry play an important role in this construction, as the extremal 
surface tends to wrap around them. 

In the main part of the paper we analyzed the form of the radial null geodesics in the parametrization of the
BTZ metric that we employed. We paid particular attention to the identification of surfaces on which the expansion of the
geodesics vanishes. We saw that the parametrization of the metric in terms of Fefferman-Graham coordinates 
covers certain parts of the full Penrose diagram. The parametrization with a static boundary covers the asymptotic regions outside
the two event horizons. The constant-time surface forms a throat, or Einstein-Rosen bridge, connecting these regions.
In the parametrization which generates a boundary scale factor corresponding to decelerating expansion, the throat extends
behind the event horizons and is time-dependent. When the expansion becomes accelerating the throat approaches the center of
the BTZ black hole. In the same time, one of the two event horizons moves to infinity, so that the parametrization covers 
again two regions of the Penrose diagram: one outside and one inside the event horizon. 

The expansion of the radial null geodesics in general may vanish on the event horizons and the throat. In particular, 
both ingoing and outgoing geodesics have zero expansion on the throat, which has an area smaller than that of the
event horizons, as it is located behind them. This indicates that this surface, despite not forming a true horizon, as we 
explained in the discussion of fig. \ref{all},  may play an important role in the 
determination of the CFT entanglement entropy through holographic methods \cite{ryu,takayanagi,hubeny}.
This issue forms part of an ongoing study.


\section*{Acknowledgments}
We wish to thank T. Christodoulakis and P. Terzis
for useful discussions. 
N.~T. was supported in part by the EU Marie Curie Network ``UniverseNet'' 
(MRTN--CT--2006--035863) and the ITN network
``UNILHC'' (PITN-GA-2009-237920).


\end{document}